\documentclass[12pt]{iopart}

\usepackage{graphicx}
\usepackage{epstopdf}
\usepackage{color}
\usepackage{amssymb}
\expandafter\let\csname equation*\endcsname\relax
\expandafter\let\csname endequation*\endcsname\relax
\usepackage{amsmath}
\usepackage{siunitx}
\usepackage{subfigure}

\usepackage{iopams}  
\begin{document}

\title[]{Birefringence Measurements on Crystalline Silicon}

\author{Christoph Kr\"uger$^1$, Daniel Heinert$^2$, Alexander Khalaidovski$^1$, Jessica Steinlechner$^3$, Ronny Nawrodt$^2$, Roman Schnabel$^{1,4}$ and Harald L\"uck$^1$}

\address{$^1$ Institut f\"ur Gravitationsphysik, Leibniz Universit\"at Hannover and Max-Planck-Institut f\"ur
Gravitationsphysik (Albert-Einstein-Institut), Callinstr 38, D-30167 Hannover, Germany}
\address{$^2$ Institut f\"ur Festk\"orperphysik, Friedrich-Schiller-Universit\"at Jena, Helmholtzweg 5, D-07743 Jena, Germany}
\address{$^3$ SUPA, School of Physics and Astronomy, University of Glasgow, \\Glasgow G12 8QQ, Scotland}
\address{$^4$ Institut f\"ur Laserphysik und Zentrum f\"ur Optische Quantentechnologien,
Universit\"at Hamburg, Luruper Chaussee 149, 22761 Hamburg, Germany}
\ead{christoph.krueger@aei.mpg.de}


\begin{abstract}
Crystalline silicon has been proposed as a new test mass material in third generation gravitational wave detectors such as the Einstein Telescope (ET). 
Birefringence can reduce the interferometric contrast and can produce dynamical disturbances in interferometers.
In this work we use the method of polarisation-dependent resonance-frequency analysis of Fabry-Perot-cavities containing silicon as a birefringent medium.
Our measurements show a birefringence of silicon along the (111) axis of the order of $\Delta\,n \approx 10^{-7}$ at a laser wavelength of 1550\,nm and room temperature.
A model is presented that explains the results of different settings of our measurements as a superposition of elastic strains caused by external stresses in the sample and plastic strains possibly generated during the production process. 
An application of our theory on the proposed ET test mass geometry suggests no critical effect on birefringence due to elastic strains.
\\
\\
DCC number: LIGO-P1500040-V1
\end{abstract}

\section{Introduction}
The initial as well as the advanced version of gravitational wave (GW) detectors, which are currently being installed, such as aLIGO and advanced Virgo, use suspended fused silica optics as test masses \cite{ligo,virgo}. These observatories are set up as dual-recycled cavity-enhanced Michelson-type laser interferometers with a kilometer-scale baseline and are operated at room temperature.
The KAGRA observatory~\cite{kagra} and parts of the proposed Einstein Telescope (ET) \cite{ETDS} will be operated at cryogenic temperatures and will use new test mass materials in order to reduce thermal noise. 
In the case of the low frequency interferometer of the Einstein Telescope (ET-LF)~\cite{ETDS} crystalline silicon has been suggested as test mass material. The test masses of GW detectors need to have low mechanical loss~\cite{nawrodt08} to limit thermal noise and low optical loss at the laser wavelength to avoid the formation of excessive thermal lenses and an overall heating of test masses. The optical absorption of silicon at the designated ET wavelength of 1550\,nm is currently being investigated~\cite{khal2013,degallaix2013,degallaix2014,glasgow}.
The test masses also need to have a low birefringence to allow a high interferometer contrast and high power- and signal-recycling gains~\cite{PR, SR, 127dB, Birefringence-ifo}.

In this paper we investigate birefringence and its effects in silicon test masses at room temperature. 
The effect of birefringence and an upper birefringence limit for GW detectors is discussed in the next section.
We use a Fabry-Perot-cavity to detect the birefringence of silicon which allows the sensing of the integrated birefringence along the axis of the cavity mode. An overview of the measurement method is given in section 2.
We present experimental results for the birefringence in silicon samples and a model reproducing the observed behavior.
Based on our results we estimate the expected birefringence in the test-masses of ET.

\subsection{Effect of birefringence in a Michelson-type interferometer}\label{effect}
\begin{figure}[h]
	\centering
		\includegraphics[width=0.9\textwidth]{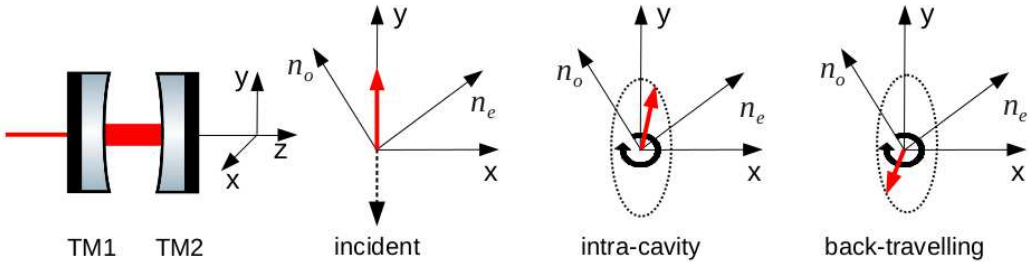}
	\caption[]{Effect of birefringence on light coupled to and reflected off an arm cavity formed by the test mass mirrors TM1 and TM2. The incident light field (red arrow) is linearly polarized along the y-axis. The axes of the ordinary and extraordinary polarizations ($n_o$ and $n_e$, respectively) do generally not align with the polarization of the incident light field. While transmitting through the incoupling test mass (TM1) the light field experiences birefringence which causes the intra-cavity light field to be elliptically polarized. The red arrow marking the polarization now circles on the dashed ellipse. 
When leaving the cavity the present birefringence will not reverse the effects of the first transmission which means the back-traveling light field is still elliptically polarized.
\label{f:cavity}}
\end{figure}

In GW detectors, linearly polarized laser light is coupled into a resonant arm cavity formed by the test masses of a GW detector (designated TM1 and TM2, respectively), see Figure \ref{f:cavity}. Due to the orthogonal axes of the indices of refraction ($n_o$ and $n_e$, respectively) which are generally not precisely aligned with the polarization of the incident light, the light field is split into two components which sense different optical path lengths inside the test mass. The two orthogonal fields transmitted through the test mass will then be out of phase, resulting in elliptically polarized light inside the cavity (Figure \ref{f:cavity}). 
In this work, birefringence-free coatings are assumed. This assumption will be justified in detail in section 4.3 (see also \cite{Moriwaki, Brandi, Camp}). In this case both polarizations experience the same phase shift under reflection at the cavity mirrors, hence no further change in polarization inside the cavity is added.
When leaving the cavity through the incoupling mirror TM1, the birefringence will (in general) further increase the ellipticity (Figure \ref{f:cavity}). When superimposed on the beam splitter of a GW detector, the two light fields emerging from the two arms  (generally having experienced different levels of birefringence) show different polarizations and lead to a reduced contrast of the interferometer. This is of concern as GW detectors are operated close to their dark fringe \cite{ETDS, dark-fringe, dark-fringe-LIGO}, which is the typical procedure to reduce the light power on the photodetector and to enable power recycling \cite{dark-fringe}.

Furthermore, each polarization dependent optical component implemented in a GW detector, such as optical isolators or polarization-dependent beam splitters, will give rise to additional optical losses once the incident polarization is modified. These losses would reduce the efficiency of advanced techniques such as the use of squeezed light \cite{sqz1, sqz2}.
If the orientation of the birefringence of the test mass materials does not depend on the location of beam transmission through the test masses, an easy solution would be to align the beam polarization and the test masses in a way such that the incident polarization is aligned with one of the orthogonal components of the index of refraction. In this case the effects of birefringence could be suppressed. 
In practice, however, such an alignment may be not perfect, and furthermore, the birefringence orientation may be a function of the position inside the mirror substrate.

In addition, dynamic effects caused by birefringence, have to be considered; for example an oscillatory rotation of the test mass around the beam axis. 
If in this case no birefringence is present, there will be no modulation of the light field. 
But, once there are two different indices of refraction, there will be an oscillatory coupling to the other polarization generating an oscillating intensity at the detection port of the interferometer, which cannot be distinguished from a gravitational wave signal.

The birefringence of fused silica test masses has been estimated to be between $2.5 \cdot 10^{-8} \leq \Delta n \leq 5.0 \cdot 10^{-8}$ \cite{Birefringence-ifo}. These values were sufficiently low for the initial generation of GW detectors. 
In Ref.~\cite{Birefringence-ifo}, Winkler \textit{et al.} gave a detailed description on how birefringence can limit a power recycled GW detector with arm cavities. Let $\frac{\Delta P}{P_0}$ be the power losses due to depolarization in a GW detector, then in order to not limit the power build-up in the interferometer $$\frac{\Delta P}{P_0} < \frac{1}{G}$$ has to be fulfilled, where G is the power-recycling gain of the interferometer.

Making further use of the equations given in~\cite{Birefringence-ifo}, we can derive an upper limit for the acceptable birefringence assuming that the axes of the indices of refraction (designated $n_e$ and $n_o$ in Figure \ref{f:cavity}) of silicon are oriented at an angle of $\pi/4$ to the linear polarization of the laser light used in the GW detector. An angle of $\pi/4$ is the worst case since it maximizes the resulting ellipticity. 
According to~\cite{Birefringence-ifo}
\begin{equation}
\frac{\lambda}{\pi L} \arcsin\left(\frac{1}{\sqrt{G}}\right) > \Delta n,
\label{e:depol-loss}
\end{equation}
with L being the substrate length and $\lambda$ being the laser wavelength. The ET-LF test masses are planned to be cylindrical silicon substrates having a thickness of about 50\,cm and a diameter of at least 45\,cm. Using parameters provided in the ET Design Study \cite{ETDS} ($\lambda = 1550$\,nm, $G=21.6$ and two transmissions through the 50\,cm silicon optics) yields $\Delta n < 10^{-7}$ as an upper limit for the tolerable birefringence.
Please note that this derivation neither takes into account signal-recycling nor the injection of squeezed states. Signal-recycling uses an additional mirror, called Signal Recycling Mirror (SRM), placed between the beam splitter and the photo diode to resonantly influence the amplitude of the signal sidebands created by GWs. The SRM reﬂects a fraction of the light, leaving the interferometer through the detection port, back into the interferometer \cite{Signalrecycling}. This conﬁguration increases the signal strength of the interferometer in a frequency range depending on the microscopic position of SRM. Generally two ways of operation are distinguished: ``Signal Recycling" where the storage time of sidebands is increased by setting the SRM position to resonance for a certain frequency, which increases the sensitivity within the bandwidth of resonance, and ``Resonant Sideband Extraction" where SRM is tuned to anti-resonance reducing the storage time of the GW sidebands in the arm cavities and thus widening the sensitive bandwidth of the interferometer. The introduction of SRM generally changes the tolerable optical losses. The ET-LF interferometer will use signal-recycling tuned to a resonance of 25\,Hz. With the ET-LF parameters the use of signal recycling does not increase the requirements for optical losses beyond the demands for power recycling.

Squeezed light injected through the signal port of the interferometer results in a more demanding limit for the maximally tolerable losses. 
The sensitivity gain in a GW detector using squeezed light depends on its optical losses.
Aiming for a squeezing level of 10\,dB requires a system with total optical losses $\Lambda < 10\%$ \cite{127dB} and hence the losses resulting from birefringence effects alone should be considerably smaller; let us assume $\Lambda < 1\%$ as a limit.
With the design parameters of ET-LF this results in $\Delta\,n < 10^{-8}$.

Relation (\ref{e:depol-loss}), however, assumes the worst case scenario regarding the angle between either of the axes of the indices of refraction and the polarization of the laser light, i.e. an angle of $\theta = 45^{\circ}$. Carefully adjusting this angle $\theta$ and assuming a constant orientation and value of birefringence in the beam volume inside the optics, can significantly relax the derived limit for the maximally tolerable birefringence.
Taking a variable misalignment angle $\theta$ into account relation (\ref{e:depol-loss}) becomes (\cite{Birefringence-ifo} and references therein)
\begin{equation}
\frac{\lambda}{\pi L} \arcsin\left(\frac{1}{\sqrt{G \sin^{2}(2\theta)}}\right) > \Delta\,n.
\label{e:depol-misalignment}
\end{equation}
By reducing the angle $\theta$ from $45^{\circ}$ to a value of $\theta = 4^{\circ}$ the tolerable birefringence increases to 
\begin{equation}
\Delta\,n\left(\theta={4^{\circ}}\right) < 10^{-7}
\label{e:depol-4deg}
\end{equation}
which is in the same order of magnitude as the initially derived limit for a power recycling gain of $G = 21.6$. If a misalignment of $\theta = 1^{\circ}$ can be achieved, the birefringence limit increases to $\Delta\,n\left(\theta={1^{\circ}}\right)< 5 \times 10^{-7}.$

\subsection{Previous studies on birefringence in silicon}
Previous measurements of birefringence of silicon have yielded variable values, some surpassing the threshold given in the previous section and others being well below that limit.
In 1959, Lederhandler examined the birefringence of silicon parallel to the $<111>$ direction \cite{Lederhandler}. The team used light with a wavelength
between $1100$\,nm  and $1200$\,nm and samples with a specific resistivity between $0.01\,\Omega$cm and $2\,$k$\Omega$cm. The measured values of the birefringence varied between $7 \times 10^{-4} < \Delta n < 9 \times 10^{-4}$.
In 1971 Pastrnak and Vedam \cite{Pastrnak} observed a birefringence of $\Delta n = 5 \times 10^{-6}$ for a wavelength of 1150\,nm in the $\left< 110 \right>$ direction and none in the $\left< 111 \right>$ and $\left< 100 \right>$ directions.
In 2001, Fukuzawa \textit{et al.} \cite{Fukuzawa} found that ``thermal processing'' of \SI{3}{\arcsecond} $(100)$ silicon wafers introduces ``anomalous'' birefringence of about $\Delta n \approx 10^{-5}$ at a light wavelength of 1300\,nm. This result points out that birefringence in silicon might have origins in the production process.
About a year later, Chu \textit{et al.} \cite{Chu} measured the birefringence of a silicon single crystal without lattice dislocations (dislocation-free). Along the $\left< 110 \right>$ configuration they measured $\Delta n = 3.2 \times 10^{-6}$ at a wavelength of 1520\,nm. Furthermore, they reported to have observed an ``extremely small'' level of birefringence when transmitting light along the $\left< 001 \right>$ direction, while their method has been capable of measuring values of the order of $\Delta n \approx 10^{-8}$. 

A summary of the results is given in Table~\ref{t-overview}. All measurements were performed with rather thin samples compared to the dimensions envisioned in ET, and furthermore lead to partially contradictory results. 
\begin{table}[h!]
\begin{center}
\begin{tabular}{|l|l|l|l|l|}
	\hline
	Direction of & Wavelength & Birefringence & Year & Source\\
	light propagation & in nm & in $\Delta n$ & & \\
	\hline
	\hline
	$\left<111\right>$ & 1100 - 1200 & $7 - 9 \times 10^{-4}$ & 1959 & Lederhandler \textit{et al.} \cite{Lederhandler}\\
	\hline
	$\left<100\right>$ & 1150 & $< 10^{-6}$ (*) & 1971 & Pastrnak \textit{et al.} \cite{Pastrnak}\\
	$\left<110\right>$ & 1150 & $5 \times 10^{-6}$ & 1971 & Pastrnak \textit{et al.} \cite{Pastrnak}\\
	$\left<111\right>$ & 1150 & $< 10^{-6}$ (*) & 1971 & Pastrnak \textit{et al.} \cite{Pastrnak}\\
	\hline
	$\left<100\right>$ & 1300 & $\approx 10^{-5}$ & 2001 & Fukuzawa \textit{et al.} \cite{Fukuzawa}\\
	\hline
	$\left<110\right>$ & 1520 & $3.2 \times 10^{-6}$ & 2002 & Chu \textit{et al.} \cite{Chu}\\
	$\left<001\right>$ & 1520 & $\approx 10^{-8}$ & 2002 & Chu \textit{et al.} \cite{Chu}\\
	\hline

\end{tabular}
\end{center}
\caption{Birefringence of crystalline silicon depending on crystal orientation and wavelength as measured in previous experiments by other authors. (*) value reported ``not observed'' in the original publication \label{t-overview}}
\end{table}
\section{Experiment}
In the scope of this work, detailed measurements were performed on sample 1, which is a 1.2\,kg test mass having a specific resistivity of 11\,k$\Omega$cm with a direction of light propagation being parallel to the $\left<111\right>$ crystal orientation. 
The data shown in Figure \ref{f:rotation} and Figure \ref{f:load} have been obtained using this sample. Furthermore, the data obtained with this sample have been used for the stress simulations presented in section 4.

For comparison and to overcome the FSR ambiguity of our measurement technique we analyzed other samples (samples 2-4) with the same crystal orientation but different thicknesses and different values of resistivity. The laser wavelength was always 1550\,nm.

Commonly, birefringence is measured by using a set of two polarizers with an angle of 90$^\circ{}$ between their optical axes. This setup prevents light from being transmitted through both polarizers as the second polarizer blocks all light which is transmitted by the first polarizer. A birefringent sample between the polarizers will partially convert the linearly polarized light transmitted through the first polarizer into the orthogonal polarization which will pass through the second polarizer. The implementation of a polarization modulator then allows a quantitative determination of the bireffringence \cite{Saph-Birefringence, PEM, PEM2}.
In this work, we use a different approach to measure birefringence, which circumvents the use of polarization modulation and hence avoids the necessity to detect small intensity variations of a light field. The method is explained in detail in the next section.

\subsection{Experimental Setup}
\begin{figure}[h!]
	\centering
		\includegraphics[width=0.9\textwidth]{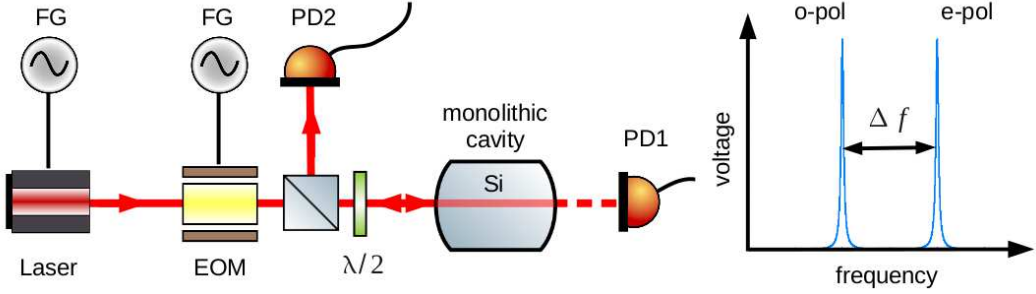}
	\caption[]{
Schematic experimental setup. The frequency of linearly polarized laser light with wavelength of 1550\,nm is varied over a range of 60\,MHz, actuating the laser's piezo-electric crystal. The light is transmitted through an electro-optic modulator (EOM) driven at 15\,MHz. The EOM is used to imprint sidebands to the light field in order to calibrate the frequency axis of the recorded data. A $\lambda/2$ wave-plate is used to adjust the polarization and hence the power in the resonant modes such that both yield comparable signal strengths. The light field transmitted through the cavity and detected by a photo detector (PD1) is used to record the intensity maxima of the cavity modes.
The reflected field (detected by PD2) is used to generate Pound-Drever-Hall type error signals which show a higher signal to noise ratio with respect to the dark noise of the photo detectors than the sidebands transmitted through the cavity. FG: function generator.\label{f:schematic}}
\end{figure}

The experimental setup is shown in Figure \ref{f:schematic}. The silicon sample under investigation is equipped with convex polished surfaces and highly reflective Ta$_{2}$O$_{5}$/SiO$_{2}$ coatings which let the sample form a monolithic cavity. A laser beam of 1550\,nm wavelength is coupled into the monolithic silicon cavity. In order to be coupled resonantly into an optical cavity the light field needs to fulfill the condition \cite{Hecht-Optics}
\begin{equation}
N \lambda = 2 n d,
\label{e:res-con}
\end{equation}
where N is an integer, $\lambda$ denotes the wavelength, n the index of refraction and $d$ is the geometric length of the cavity.

Birefringence means two different indices of refraction $n_e$ and $n_o$ for light of two orthogonal linear polarizations (designated e-pol and o-pol in Fig~\ref{f:schematic}, respectively).
Tuning the laser frequency $\nu$ produces two resonances per free spectral range - one per polarization - that can be detected in transmission and reflection of the cavity. The frequency spacing $\Delta\,f$ of the resonances allows the deduction of the level of birefringence of the silicon sample under investigation. Similar approaches of measuring birefringence have been used earlier in order to measure the birefringence of high reflective mirrors \cite{Moriwaki,Brandi}

In our experiment, the laser was phase modulated with a frequency $\nu_m \ll \nu$. The resulting sidebands served as calibration markers for measuring the frequency difference $\Delta\,f$ of the two birefringent cavity modes. The laser frequency actuator itself was not sufficiently linear over a full free spectral range of the cavity.

\subsection{Theoretical Background}
Assuming a birefringent cavity as described above. Let $n_o$ and $n_e$ denote its orthogonal indices of refraction.
Similarly, let $\nu_o$ and $\nu_e$ denote the laser frequencies, which let $\lambda_o$ and $\lambda_e$ meet the resonance condition (Equation \ref{e:res-con}) with a given number of nodes $N_o$ and $N_e$ inside the cavity, respectively.
Having defined the frequency spacing between the resonance maxima as $\Delta\,f$ and denoting the birefringence $\Delta\,n$, we get $\nu_e = \nu_o-\Delta\,f$ and similarly $n_e = n_o +\Delta\,n$.

The resonance conditions (Equation \ref{e:res-con}) for both polarizations then is
\begin{equation}
N_o = \frac{2d\nu_o n_o}{c_0} \textup{ and } N_e = \frac{2d\nu_e n_e}{c_0} ,
\end{equation}
with $c_0$ denoting the speed of light in vacuum. Assuming $\Delta\lambda=\lambda_e-\lambda_o \ll \lambda_o$ both polarizations are resonant with the same number of nodes inside the cavity (we will justify this assumption below) and hence $N_o = N_e$.
Setting $\nu_o = \nu$ and $n_o=n$ leads to $\nu_e = \nu - \Delta\,f$ and $n_e = n+\Delta\,n$ which allows to omission of the indices. Since these conditions are valid for the same number of nodes $N$, this leads to 
\begin{equation}
\Delta\,n = n \cdot \frac{\Delta\,f}{\nu-\Delta\,f}
\label{e:delta-n}
\end{equation}
which gives us the birefringence $\Delta\,n$ once the frequency difference $\Delta\,f$ between the two resonances of the cavity is known.

The measured frequency spacing $\Delta\,f$ is, however, ambiguous as it can only be measured modulo free spectral ranges $\mathcal{FSR} = c_0 / 2dn$ of the cavity. If $\Delta\,f$ were to be greater than one free spectral range, one polarization would fulfill the resonance condition with a given number of nodes $N$ while the perpendicular polarization would fulfill it with another number of nodes $N + \Delta\,N$, where $\Delta\,N$ is an integer. Since the mode pattern of the cavity repeats every free spectral range, it is not possible to distinguish whether two maxima arise out of the same free spectral range when simply observing the light transmitted through the cavity.

This ambiguity could be overcome by changing the laser frequency over hundreds FSRs of the cavity. If the peaks do not arise from the same FSR, a small change of $\Delta\,f$ should be observable. The expected change of $\Delta\,f$ for our largest possible laser frequency change was calculated to be lower than the accuracy of our measurement method.
This problem, however, can be circumvented by changing the FSR of a the cavity. In this work we exploit the dependence of the cavity's FSR on the cavity length $d$ and measure $\Delta\,f$ of different cavities made from the same material but having different lengths $d$.

\section{Silicon Samples}
Silicon has a cubic face-centered lattice structure \cite{HoOM}. 
It is known that such structures have vanishing natural birefringence, however, applied stress and external loads can change this behaviour \cite{HoOM}. 
As the test masses in interferometric GW detectors have weights of many kilograms and are suspended as pendulums, it has to be examined in which way stress affects the optical properties of silicon.

\subsection{Overview of samples}
Within this work a silicon sample with a mass of 1.2\,kg and cylindrical shape (with small lateral flat areas) has been examined (sample 1). Three additional samples (samples 2,3 and 4) of different thicknesses have been used to determine whether the observed modes resonate in the same FSR ($\Delta\,N = 0$).

These additional samples have thicknesses between 2.8\,cm and 9.9\,cm, have a diameter of 2.4\,cm, and give rise to free spectral ranges covering the range from 435\,MHz to 1.54\,GHz. 
The values obtained for the frequency spacing $\Delta\,f$ are at least two orders of magnitude below the respective free spectral ranges (see Table \ref{t:substrates}).
This finding strongly suggests that the observed transmission maxima indeed resonate in the same free spectral range, hence $\Delta\,N = 0$ allowing the application of the theory presented above.

\begin{table}[h!]
\begin{tabular}{|r|r|r|r|r|r|r|r|}
	\hline
	Sample & Thickness & Diameter & Resistivity & Mass & $\Delta\,f$,& $\Delta\,n \times 10^{-7}$ & FSR\\
	Number & in mm & in mm & in k$\Omega$cm & in kg & in MHz& & in MHz\\
	\hline
	\hline
	1 & 65 & 100 & 11 & 1.2 & 1.26 - 5.98 & 0.23 - 1.07 & 663\\
	\hline
	2 & 28 & 24 & 2 & 0.059 & 6.00 - 6.20 & 1.08 - 1.11 & 1539\\
	3 & 30 & 24 & 30-70 & 0.063 & 1.85 - 2.71 & 0.33 - 0.49 & 1437\\
	4 & 99 & 24 & 2 & 0.21 & 0.48 - 0.96  & 0.09 - 0.17 & 435\\
	\hline

\end{tabular}
	\caption[]{Dimensions and physical properties of silicon samples under investigation. Masses have been calculated with a density of 2330\,kg/m$^{3}$ \cite{HoOM}. All substrates have the shape of cylinders with biconvex faces with the light propagating along the (111) direction. The radius of curvature of the faces through which light is coupled into the substrate equals 2\,m for all samples except for the 65\,mm sample which has a radius of curvature of 1\,m.\label{t:substrates}}
\end{table}

The values of the birefringence for the laser beam propagating along the (111) axis are below $\Delta\,n \leq 0.49 \cdot 10^{-7}$ for samples 3 and 4.
This value of birefringence is well below the limit of $\Delta\,n < 10^{-7}$ which has been derived in section\,\ref{effect} for arbitrary test mass orientation. 
With $\Delta\,n \leq 1.11 \times 10^{-7}$ sample 2 shows a birefringence which is slightly above the derived limit. However, aligning the sample better than $\theta = 4^{\circ}$ would keep the birefringence within the derived limits of section\,\ref{effect}.

The experimentally obtained values of the birefringence for these four samples significantly scatter, strongly depending on the sample under investigation. As inherent strain has been observed in sample 1 (see section below) it seems likely that such inherent strain is present in the remaining samples as well. Such inherent strain resulting from the production process can explain the scattering of the measured data observed in this work.

\subsection{Sample 1}
With strong experimental evidence that the assumption $\Delta\,N=0$ is correct sample 1 has been used for a deeper analysis of birefringence in silicon.

\begin{figure}[h!]
	\centering
		\includegraphics[width=0.7\textwidth]{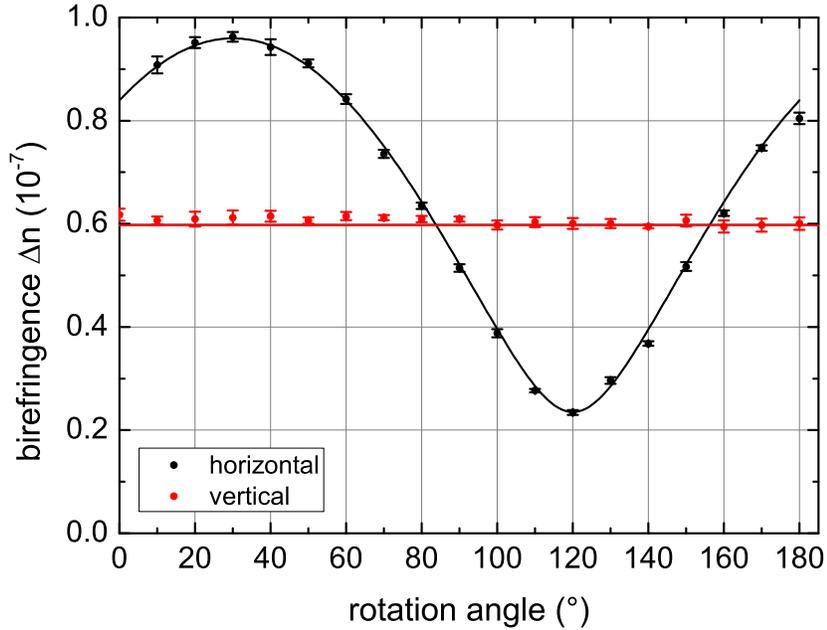}
	\caption[]{Dependence of the birefringence of sample 1 ($\varnothing$\,65\,mm$\times$100\,mm) on the angle of rotation $\phi$. If the gravitational force acts parallel to the optical axis of the cavity the birefringence stays constant, independent of $\phi$ (red data points). If the gravitational force acts perpendicular to the optical axes a clear periodic pattern is observed (black data points). The lines indicate the results of our numerical calculations.
\label{f:rotation}}
\end{figure}

\begin{figure}[ht]
\centering
\subfigure[]{
   \includegraphics[scale =.4] {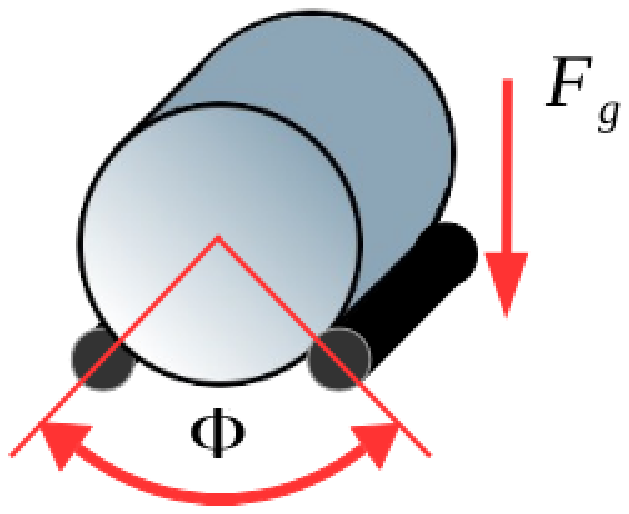}
   \label{fig:supp1}
 }
 \subfigure[]{
   \includegraphics[scale =.4] {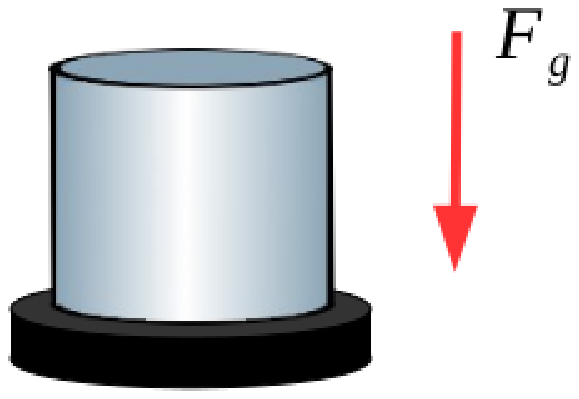}
   \label{fig:supp2}
 }
 \subfigure[]{
   \includegraphics[scale =.4] {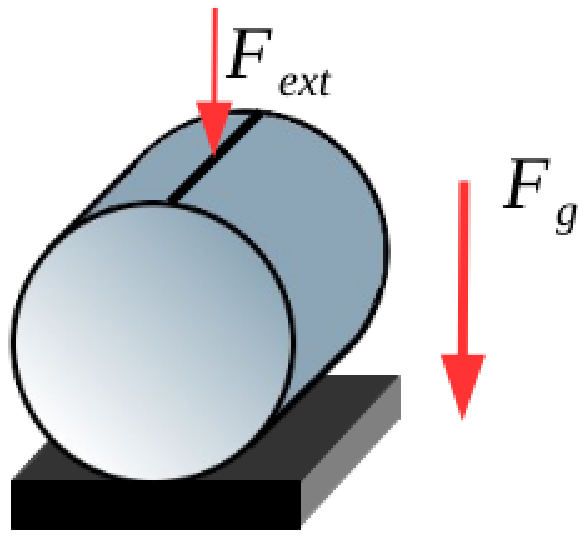}
   \label{fig:supp3}
 }
\subfigure[]{
   \includegraphics[scale =.4] {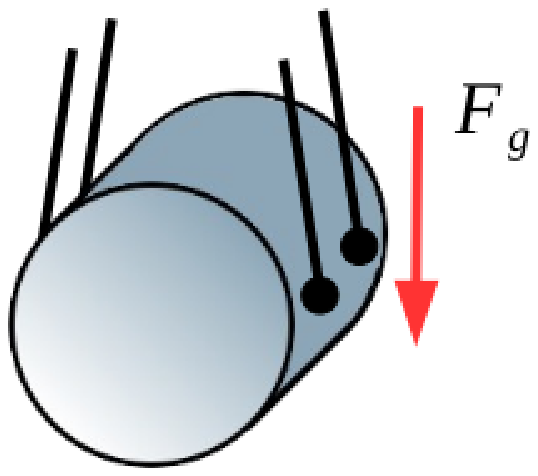}
   \label{fig:supp4}
 }
\label{myfigure}
\caption{Schematic of different ways to support the samples under investigation. The sample is colored blue while the support structures are black. a) Support used for samples 1-4 in the scope of this work. The sample is supported by two different Teflon bars forming the angle $\phi$. b) Sample 1 has been placed in a way such that the gravitational force acted parallel to the cylinder axis, thus eliminating any effects perpendicular to the direction of beam propagation. c) Support used to determine the dependence of birefringence of the external forces. The sample is supported by a single bar directly from beneath. The load is placed on top. d) Test mass suspension of GW detectors using wires attached to the sample. }
\end{figure}

In order to determine the effects of the test mass weight on the birefringence two different experiments have been performed. First the sample has been placed ``vertically" such that the gravitational force acts parallel to the optical axis of the cavity (see Fig.~\ref{fig:supp2}). 
In this position a birefringence of $\Delta\,n = 0.61\times 10^{-7}$ has been observed, independent of a rotation around the sample's cylindrical axis as depicted in Figure \ref{f:rotation}. 
Due to the problem's symmetry the gravitational force causes a radially symmetric strain inside the sample leading to a vanishing birefringence along the cylindrical axis.
Hence no birefringence ($\Delta\,n = 0$) is expected, which is in contrast to the experimental finding.

In a second measurement the cavity has been placed ``horizontally" such that the gravitational force is acting perpendicular to the optical axis. 
In this configuration, which is shown in Fig.~\ref{fig:supp1} with an angle between the supporting rods of $\phi=\SI{120}{\degree}$ (see figure caption), the birefringence has been measured at different angles $\alpha$ representing a rotation along the optical axis of the cavity.
In this case the gravitational force should lead to an anisotropic stress configuration giving rise to birefringence. But the symmetry of such a rotation predicts a constant level of birefringence.
In contrast to these considerations the experimental results of birefringence show a periodic pattern presented in Fig.~\ref{f:rotation}.
The results range between $0.23 \leq \Delta\,n \times 10^{7} \leq 1.07$.
The mean value of $\Delta n$ is close to the birefringence, which has been observed when the gravitational force acts parallel to the optical axis of the sample.
This superposition of intrinsic and external stress would allow a minimization of the overall birefringence if the required external stress could be produced by the test mass suspension in a GW detector and the orientation and level of intrinsic birefringence was predictable.

\begin{figure}[h!]
	\centering
		\includegraphics[width=0.7\textwidth]{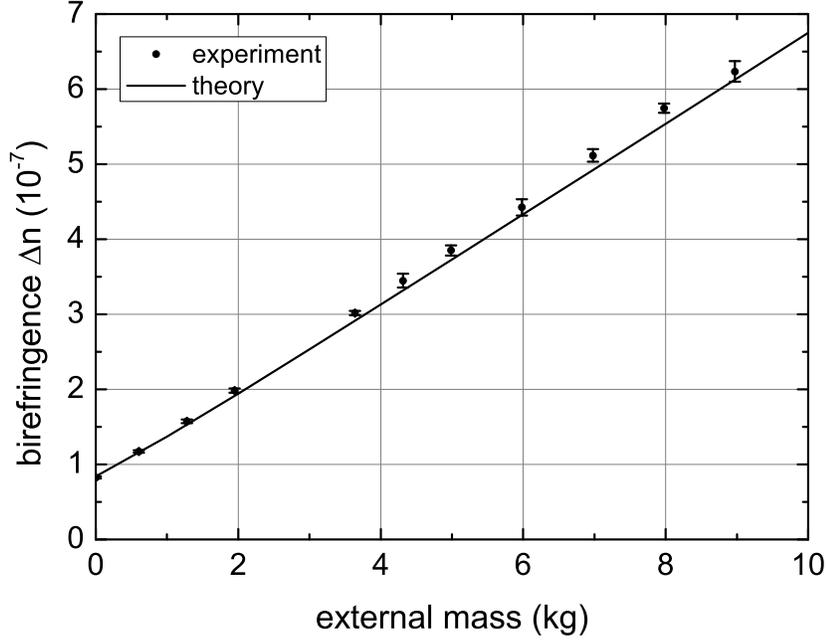}
	\caption[]{Dependence of the birefringence of sample 1 ($\varnothing$\,65\,mm$\times$100\,mm) on the weight of an external load (elastic strain). A clear linear dependence between the measured birefringence and a mass placed on top of the sample as shown in Fig.~\ref{fig:supp3} has been observed.
\label{f:load}}
\end{figure}

In order to determine how external forces might change the measured birefringence, the sample has been placed onto one of its flats while an extra weight has been placed on top of the opposing flat (see Fig.~\ref{fig:supp3}). A clear linear dependence between the birefringence $\Delta\,n$ and the external load has been observed as shown in Figure~\ref{f:load}. 
Up to an external weight of 9\,kg, no deviation from the linear behaviour has been observed.

The behaviour of the birefringence is, however, subject to the respective mounting. 
While a load dependence has been seen using a support from below (Fig.~\ref{fig:supp3}) a load independent birefringence is obtained using two supports at $\phi=\SI{120}{\degree}$ shown in Fig.~\ref{fig:supp1} and an external load on top.
Due to the symmetry of the support in the latter setup an external load from the top of the sample is not expected to cause an anisotropic stress along the axis of the test mass. Thus a change of the external load should not change the observed level of birefringence.

\section{Stress Simulations}
Figures~\ref{f:rotation} and \ref{f:load} show that external forces acting on sample 1 cause additional birefringence inside the crystal. Furthermore, birefringence is present even if no external force is applied to the sample. In order to explain this behavior and to predict the effect of birefringence considering the test masses and their suspensions as proposed for the ET design, a finite element analysis (FEA) was conducted.

\subsection{Relation of Strain and Birefringence}
Any mechanical force acting on a solid will cause stress and strain within the sample.
Via the photoelastic effect the strain $u_{kl}$ in the sample causes a change of the optical properties that is described by the tensor of impermeability $B_{ij}$.
Using the photoelastic tensor $p_{ijkl}$ the change of the tensor of impermeability can be expressed as
\begin{equation}
	\Delta\,B_{ij} = p_{ijkl} u_{kl} \ .
	\label{e:stress-impermeability}
\end{equation}
In the following we exclusively use the Voigt contraction of index pairs ($1\leftrightarrow xx$, $2\leftrightarrow yy$, $6\leftrightarrow xy$) yielding
\begin{equation}
	\Delta\,B_{i} = p_{ij} u_{j} \ .
\end{equation}

In the strainless state silicon with its cubic lattice shows isotropic optical properties.
Thus, the ellipsoid representing the tensor of impermeability shows a spherical shape.
Any strain inside the sample will deform this sphere to an ellipsoid.
An incoming wave will cut this ellipsoid by the plane normal to its wave vector resulting in a two-dimensional ellipse.
The two semiaxes of this ellipse then determine the effective difference in the refractive index of differently polarized light.
Adjusting the wave vector of the incoming wave along the $z$ axis of the material coordinate system the ellipsoidal equation reads
\begin{align}
B_1 x^2+B_2y^2+2B_6 xy=1 \ .
\end{align}
To find the semiaxes of this ellipse a rotation of the coordinate system by an angle $\varphi$ with
\begin{align}
\tan(2\varphi)=\frac{2B_6}{B_1-B_2} \ ,
\end{align}
has to be applied.
In this new coordinate system the new coefficient $\tilde{B}_6$ vanishes and the semiaxes are obtained from the remaining parameters as
\begin{align}
\tilde{B}_1=B_1\cos^2\varphi+B_2\sin^2\varphi+2B_6\cos\varphi\sin\varphi \ , \\
\tilde{B}_2=B_1\sin^2\varphi+B_2\cos^2\varphi-2B_6\cos\varphi\sin\varphi \ .
\end{align}
These are connected to a change in the refractive index $\Delta n_i$ via
\begin{align}
\tilde{B}_i=\frac{1}{(n_0+\Delta n_i)^2} \ .
\end{align}
In the approximation of small strains and thus small changes in the impermeability an expression for the difference in the refractive indices due to birefringence can be found.
It reads
\begin{align}
\Delta n=\Delta n_1-\Delta n_2=-\frac{n_0^3}{2}(\tilde{B}_1-\tilde{B}_2) \ ,
\label{equ:Deltan}
\end{align}
where $n_0$ represents the isotropic refractive index of unstrained silicon.

\subsection{Plastic strains}
The stress induced birefringence effects discussed above are not sufficient to completely describe the experimental results.
Firstly, the vertically aligned cavity (Fig.~\ref{fig:supp2}) reveals a non-vanishing birefringence.
Due to the rotational symmetry of the gravitational load and the elastic properties of the crystal birefringence should vanish along the cylindrical axis.
Secondly, the rotation results of the horizontally aligned cavity shows a non-vanishing mean as expected due to gravitational strains.
However, in the experiment this constant value is superimposed by an additional modulation of birefringence exhibiting a two-fold symmetry.
As the elastic constants of silicon show a three-fold symmetry along its crystalline $\left<111\right>$ axis, the above behavior cannot be explained by the rotation of the elasticity matrix.

These experimental findings lead us to introduce another strain contribution causing birefringence in our sample.
In a simple model we allowed for an additional {\it plastic strain\/} whose orientation is fixed within the sample.
Assuming a uniaxial character of such a plastic strain results in the following Voigt notation
\begin{align}
u^{p}=(u_0,0,0,0,0,0) \ .
\end{align}
A rotation of the sample can be considered as a rotation of the coordinate system to describe the plastic strain tensor introduced above.
Following the tensor laws a rotation of the coordinate system by an angle $\alpha$ leads to modified strain coefficients of
\begin{align}
u^{p}=(u_0 \cos^2\alpha,u_0 \sin^2\alpha,0,0,0,-2u_0 \sin\alpha\cos\alpha ) \ .
\label{equ:plasticstrain}
\end{align}

Plastic strains are known to be produced in the manufacturing process of crystalline silicon samples.
Mainly during the cooling process temperature gradients arise within the crystal leading to thermal strains.
These thermal strains are frozen in during the cooldown process and remain as plastic strains once the sample reaches a homogeneous temperature distribution at room temperature.
See e.\,g. Ref.~\cite{Lederhandler} for a more detailed discussion of this process.

\subsection{Simulation}

Considering the total strain within the sample as the sum of elastic strains $u^{el}$ due to gravitational and external loads and plastic strain $u^{p}$ 
\begin{align}
u^{tot} = u^{el} + u^{p} \ ,
\label{equ:epstot}
\end{align}
allows an efficient explanation of the experimental results.
Above the elastic strains $u^{el}$ have been obtained from a 2D plane strain analysis using the finite element package COMSOL.
In this calculation we kept the global coordinate system fixed and accounted for the rotation of the sample by the modification of the coefficient of the elasticity tensor.
The plastic contribution has been taken from Eq.~(\ref{equ:plasticstrain}).
Finally inserting $u^{tot}$ from Eq.~(\ref{equ:epstot}) into Eq.~(\ref{e:stress-impermeability}) yields the change of the impermeability tensor.
From this the birefringence can be calculated via (\ref{equ:Deltan}).
Please note that for these calculations literature values for the tensor of elasticity as well as for the photoelastic tensor have been used.
Due to its cubic structure the Voigt notation of the tensor of elasticity for silicon reads
\begin{align}
C_{ij}=
\begin{pmatrix}
c_{11} & c_{12} & c_{12} & 0 & 0& 0 \\
c_{12} & c_{11} & c_{12} & 0 & 0& 0 \\
c_{12} & c_{12} & c_{11} & 0 & 0& 0 \\
0 & 0 & 0 & c_{44} & 0& 0 \\
0 & 0 & 0 & 0 & c_{44}& 0 \\
0 & 0 & 0 & 0 & 0& c_{44} \\
\end{pmatrix} \ .
\label{equ:shapeC}
\end{align} 
In the following we use the values $c_{11}=\SI{165.7}{\giga\pascal}$, $c_{12}=\SI{63.9}{\giga\pascal}$ and $c_{44}=\SI{79.6}{\giga\pascal}$ from Ref.~\cite{mcskimin1953}.
While in general the tensor of photoelasticity is not symmetric the point group of silicon results in the same matrix structure as for the tensor of elasticity.
Following Biegelsen \cite{biegelsen1974} we use $p_{11}=\SI{-0.094}{\giga\pascal}$, $p_{12}=\SI{0.017}{\giga\pascal}$ and $p_{44}=\SI{-0.051}{\giga\pascal}$ as photoelastic coefficients in silicon.
As we fix the global coordinate system in our analysis the coefficients of the material tensors have to be adopted due to rotation.
For completeness we state this rotated form in \ref{app:rotatedtensor}. 

With this model considering plastic as well as elastic strains in the sample we are able to fully explain the behavior of a horizontally and a vertically aligned cavity as presented in Fig.~\ref{f:rotation}.
In this respect the theoretical curves for all measurements have been obtained with a single fitting parameter for the plastic strain $u_0=\SI{2.7e-8}{}$ that is characteristic to our sample.
It turns out that in our sample the birefringent effects of plastic strains and strains induced by gravitation are in the same order of magnitude.
Further, the birefringence with respect to an additional external loading is explained by the same approach which can be seen in Fig.~\ref{f:load}.

Birefringence in the samples examined within the scope of this work can be caused by the bulk material or by the coated cavity surfaces. The latter is realized as a classical $\lambda/4$ layer stack of two alternating coating materials exhibiting a high and low index of refraction. Although the stack shows 20 layer pairs in total most of the light is reflected by the first few coating layers, significantly reducing the effective coating thickness. The substrate has, however, a thickness which is about four orders of magnitude larger than the entire coating material. Consequently, the coating had to cause a birefringence $10^4$ times bigger than the birefringence of the bulk material in order to cause the same frequency spacing. Furthermore, the application of the coating is performed at significantly lower temperatures than the production of silicon. For this reason the frozen-in plastic strains are expected to be much smaller in the coating as well. Moreover, the birefringence induced in the coating due to differences in the coefficients of thermal expansion between bulk and coating should lead to a radially symmetric pattern with no strain, and consequently negligeable birefringence, on the symmetry axis. We hence neglected the contribution of the coating to the birefringence in the analysis above an approach which is supported by the values given in \cite{Moriwaki, Brandi, Camp}.

In our model plastic strains are fixed in the sample's coordinate system while the elastic strain is oriented along the gravitational force and thus fixed in the global coordinate system.
Rotating the sample around its cylindrical axis thus leads to a variation of the total strain in the sample.
This allows the minimization of the birefringent effects due to a partial compensation of both strains by a respective choice of the sample's rotation angle.
Such a scheme also allows for the reduction of optical losses due to birefringence in future gravitational wave detectors if gravity induced strain and internal strain are of comparable magnitude.

\subsection{Predictions for the Einstein Telescope}
In this section our experimental results are transferred to suspended pendula as shown in Fig. \ref{fig:supp4}. 
To incorporate the effect of the different mechanical load scheme we ran another FEA calculation.
We focused on predicting the birefringence introduced by gravity using the same parameters that have been successfully used to explain our experimental findings before.

\begin{figure}[h!]
	\centering
		\includegraphics[width=0.5\textwidth]{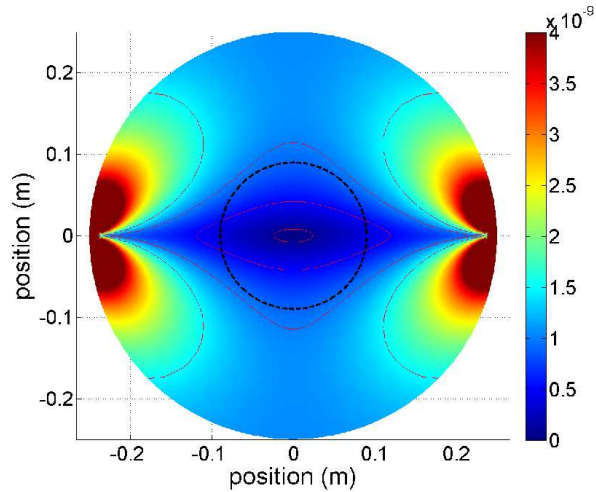}
	\caption[]{Modelled refractive index change in a suspended silicon ET-LF test mass due to stress induced birefringence due to gravity. The color code shows that high values for $\Delta n$ are concentrated around the suspension points of the sample. The red lines indicate levels of constant $\Delta n$ of \SI{0.1e-9}, \SI{0.5e-9}, \SI{1.0e-9}, and \SI{1.5e-9}. Within the laser beam radius (dashed circle) the amplitude of birefringence is well below \SI{e-9}{}. Thus, birefringence due to the gravitational load should not affect the design sensitivity of ET-LF.

\label{fig:resultsET}}
\end{figure}

This additional computation has been performed on a test mass geometry proposed for the low frequency interferometer of the Einstein Telescope (ET-LF), i.\,e. using cylindrical substrates with a diameter of 45\,cm and a thickness of 50\,cm.
For any point on the sample the difference of the index of refraction due to birefringence has been evaluated using Eq.~(\ref{equ:Deltan}) and is presented in Fig.~(\ref{fig:resultsET}).
Across the sample the values of birefringence change significantly between $\Delta\,n \approx 10^{-8}$ close to the suspension points and $\Delta\,n \leq 10^{-10}$ within the central part ($r<\SI{1}{cm}$) of the test mass. 

In contrast to our cavity measurements the beam size (ca. 86\% of power inside) in ET-LF will be about 9\,cm to reduce thermal noise.
Such a beam will sense not only a single point on the test mass but a larger area.
In order to determine the total level of birefringence encountered by a transmitting beam a sophisticated averaging has to be applied. 
For our purpose, however, it is sufficient to take the maximum value of $\Delta \,n$ within the beam diameter and use it for calculations of a worst case scenario.
Our simulation yields a maximum birefringence of $\Delta \,n \approx 10^{-9}$ in the area of beam transmission.
These values are two orders of magnitude below the upper limit of $\Delta n < 10^{-7}$ obtained earlier. 
Although in our analysis we used a 2D model including suspended lines instead of suspended points this result suggests, that elastic birefringence should be no limit for the Einstein Telescope.
However, this calculation gives no insight into the birefringence due to plastic strains, which have been causing a higher level of birefringence ($\Delta\,n \approx 10^{-7}$) in the samples examined in the scope of this work and may be even larger for ET size test masses.

\subsection{Comparison of Test Mass Materials}
Sapphire has been proposed as a test mass material which will be implemented in the KAGRA gravitational wave detector.
While silicon is isotropic, sapphire is a uniaxial material \cite{HoOM} which gives rise to an intrinsic birefringence between the c-axis and the a-axes of $|n_a-n_c| = 8\times 10^{-3}$. Letting the light propagate along the c-axes of the sapphire crystal should minimize the birefringence.
In their work Yan \textit{et al.} \cite{Yan} measured values for the birefringence in sapphire cylinders of 15 cm diameter and 6 cm thickness with light propagating parallel to the c-axis to exceed the values of silicon obtained in this work by 30-40\%.

\section{Conclusion}

In this paper we present a sensitive method for the experimental characterization of birefringence in optical materials. 
It is based on the evaluation of the frequency spacing between the two orthogonally polarized optical resonances of a monotlithic cavity made of the respective material.
Applying this method to a monocrystalline silicon (111) cavity at room temperature, we investigated the dependence of birefringence on a rotation of the sample as well as on an external load.
The method presented can be applied to cavities at arbitrary temperatures which allows the examination of silicon birefringence at cryogenic temperatures without changing the measurement method. Since the test masses of the low frequency interferometers of ET will be operated at such temperatures, the measurements performed in the scope of this work should be repeated at cryogenic temperatures.
Utilizing FEA calculations allowed us to explain the observed behavior by a superposition of elastic strains due to gravity or external loads as well as plastic strains that show a fixed orientation within the sample.
The measurements have further shown that it was possible to significantly reduce the amount of present birefringence by choosing an appropriate superposition of elastic and plastic strain in the sample. 
This is, however, only possible as long as both elastic and plastic strain are on the same order of magnitude.

Applying our results to the proposed ET-LF design suggests that the gravity-induced birefringence along the test mass axis is negligibly small in this detector.
Thus the level of plastic strains in the sample are likely to dominate the birefringence in such an application.

The injection of squeezing demands a level of birefringence which can be met by aligning sample and polarization better than 4$^{\circ}$. 
Such an alignment would allow the use of all sample materials examined in this work in a GW detector. If local variations of the level of birefringence or a locally variable orientation of the axes of the indices of refraction existed, however, the effect of this approach would be limited.
A more detailed investigation of the sources of plastic strains, including large silicon crystals, should be performed in the future.
Former works \cite{Lederhandler} on crystalline silicon already identified a clear correlation between regions showing a high density of dislocations and regions showing an increased level of birefringence.
While current silicon crystals show nearly no dislocations our experiments revealed that there is still a considerable contribution of plastic strains to birefringence.
Annealing the samples or reducing the cooling rates during and after crystal growth may be a way to minimize plastic strains.
Further experiments on birefringence distribution can help to identify the geometry of microscopical defects frozen in during crystal growth and causing plastic strains in the sample.

\section*{Acknowledgements}
We acknowledge financial support from the SFB/Transregio 7, the Aspera 3$^{rd}$ common call "ET R\&D - Networking and R\&D for the Einstein Telescope" and the EU under the "ELiTES" project (IRSES no. 295153).


\appendix
\section{Coefficients of rotated coordinate system}
\label{app:rotatedtensor}
In cubic systems the simplest shape of a fourth order tensor can be obtained by chosing the (100) axes as basis vectors ($e_x$, $e_y$, $e_z$).
Then the Voigt representation of the tensor exhibits the shape shown in Eq.~(\ref{equ:shapeC}).
To describe a cylinder in its geometry coordinate system the cylindrical axis should be along the $z$ axis of a new coordinate system ($e'_x$, $e'_y$, $e'_z$).
For the case of a (111) orientation the new base vector $e'_z$ should point along the (111) direction of the old basis, i.\,e. should be proprotional to the sum of the old basis $e_x+e_y+e_z$.
The general transformation between the two systems of basis vectors reads
\begin{align}
e'_i=a_{ij}e_j \ ,
\end{align}
with 
\begin{align}
\begin{pmatrix}
e'_x \\
e'_y \\
e'_z 
\end{pmatrix} 
=
\begin{pmatrix}
-\frac{\cos \alpha}{\sqrt{6}}+\frac{\sin \alpha}{\sqrt{2}} & -\frac{\cos \alpha}{\sqrt{6}}-\frac{\sin \alpha}{\sqrt{2}} & \frac{2 \cos\alpha}{\sqrt{6}} \\
\frac{\cos \alpha}{\sqrt{2}}+\frac{\sin \alpha}{\sqrt{6}} & -\frac{\cos \alpha}{\sqrt{2}}+\frac{\sin \alpha}{\sqrt{6}} & -\frac{2 \sin\alpha}{\sqrt{6}} \\
\frac{1}{\sqrt{3}} & \frac{1}{\sqrt{3}} &\frac{1}{\sqrt{3}} 
\end{pmatrix}
\begin{pmatrix}
e_x \\
e_y \\
e_z 
\end{pmatrix} 
\ .
\end{align}
In the equation above $\alpha$ represents the rotational angle along the cylindrical axis which remains as a degree of freedom in the orientation of the new coordinate system.
With the knowledge of the coordinate transform tensor $a_{ij}$ the coefficients of the fourth rank material tensors in new coordinates are available as
\begin{align}
C'_{ijkl}=a_{i\tilde{i}}a_{j\tilde{j}}a_{k\tilde{k}}a_{l\tilde{l}} C_{\tilde{i}\tilde{j}\tilde{k}\tilde{l}} \ .
\end{align}
An explicit calculation and transform to Voigt notation reveals
\begin{align}
C'_{ij}=
\begin{pmatrix}
c'_{11} & c'_{12} & c'_{13} & c_{14} & -c_{15}& 0 \\
 & c'_{11} & c'_{13} & -c_{14} & c_{15}& 0 \\
 &  & c_{33} & 0 & 0& 0 \\
 &  &  & c_{44} & 0& c_{15} \\
 &  &  &  & c_{44}& c_{14} \\
 &  &  &  & & c_{66} 
\end{pmatrix}
\ ,
\end{align}
with a symmetric expansion to the lower half of the matrix.
The connection to the three original coefficients is obtained by the following equations
\begin{align}
c'_{11}&=\frac{1}{2} \left(c_{11}+c_{12}+2 c_{44} \right) \, ,& \ c'_{12}&=\frac{1}{6} \left(c_{11}+5c_{12}-2 c_{44} \right) \ , \\
c'_{13}&=\frac{1}{3} \left(c_{11}+2c_{12}-2 c_{44} \right) \, ,& \ c'_{14}&=\frac{\sin(3\alpha)}{3\sqrt{2}} \left(2c_{44}+c_{12}- c_{11} \right)\ , \\
c'_{15}&=\frac{\cos(3\alpha)}{3\sqrt{2}} \left(2c_{44}+c_{12}- c_{11} \right) \, ,& \ c'_{33}&=\frac{1}{3} \left(c_{11}+2c_{12}+4 c_{44} \right)\ , \\
c'_{44}&=\frac{1}{3} \left(c_{11}-c_{12}+ c_{44} \right) \, ,& \ c'_{66}&=\frac{1}{2} \left(c'_{11}-c'_{12} \right)=\frac{1}{6} \left(c_{11}-c_{12}+4 c_{44} \right) \ . \\
\end{align}
The same rule of transformation holds for the photoelastic tensor in silicon as it exhibits the same structure as the elasticity tensor.
Further these new coefficients $c'_{ij}$ enter into the final evaluation of the birefringence.
\\
\\

\section*{References}
\bibliography{paper_V4}


\end{document}